\documentclass[a4paper,12pt]{article}

\usepackage[a4paper,left=2.73cm,right=2.7cm,top=3cm,bottom=3.5cm]{geometry}
\usepackage[colorlinks=true,linktocpage=true,linkcolor=blue,citecolor=blue]{hyperref}

\usepackage{amsmath,amssymb,graphicx}

\def\d{\mathrm{d}}

\def\le{\left}
\def\ri{\right}
\def\lab{\label}
\def\be{\begin{equation}}
\def\ee{\end{equation}}
\def\6{\partial}
\def\m{\mu}

\def\tr{{\rm Tr}}
\def\la{\langle}
\def\ra{\rangle}

\begin{document}

\begin{titlepage}

\hfill{ICCUB-14-060}

\vspace{1cm}
\begin{center}

{\LARGE{\bf Horizon universality and anomalous conductivities}}

\vskip 45pt
{\large \bf Umut G\"ursoy$^{1}$ and Javier Tarr\'\i o$^{2}$}

\vskip 20pt
{$^{1}$Institute for Theoretical Physics and Spinoza Institute,
 Utrecht University \\ 3508 TD Utrecht, The Netherlands \\}

\vskip 10pt
{$^{2}$Departament de F\'\i sica Fonamental and Institut de Ci\`encies del Cosmos, \\ Universitat de Barcelona, Mart\'\i\  i Franqu\`es 1, ES-08028, Barcelona, Spain.}\\

\vskip 10pt
{e-mails: u.gursoy@uu.nl, j.tarrio@ub.edu}
\end{center}

\vspace{10pt}
\abstract{\normalsize
We show that the value of chiral conductivities associated with anomalous transport is universal in a general class of  strongly coupled quantum field theories.  Our result applies to theories with no dynamical gluon fields and  admitting a gravitational holographic dual in the large N limit. On the gravity side the result follows from near horizon universality of the fluctuation equations, similar to the holographic calculation of the shear viscosity.       
}

\end{titlepage}

\section{Introduction} 

Anomalous transport in quantum field theories with chiral fermions has enjoyed a  renewal of interest since the recent discovery of the Chiral Magnetic Effect (CME) \cite{CME1,CME2}. In short, the CME refers to generation of a macroscopic electric current as a result of the axial anomaly in the presence of an external magnetic field $\vec{B}$ 
\be\lab{J1} 
\vec{J} = \sigma_\text{VV}\, \vec{B}  \ .
\ee
Here the ``chiral magnetic conductivity'' $\sigma_\text{VV}$ is proportional to the anomaly coefficients. The main motivation to study the CME in the context of particle physics stems from its possible realization in the Heavy Ion Collision experiments. Indeed, in an off-central collision of heavy ions at RHIC and LHC, huge magnetic fields are expected to be generated by the ``spectator'' ions that do not participate in the formation of the Quark-Gluon Plasma (QGP) \cite{MagField}. Then one can theoretically demonstrate \cite{CME1,CME2} that the chiral anomaly in QCD with electromagnetic and gluon contributions gives rise to the CME in the off-central heavy ion collisions. 

Presence of such anomaly-induced electric and chiral currents in the QGP might have imprints on the spectra of charged hadrons observed in the heavy ion collisions. Such experimental evidence is still controversial at present \cite{Exp}. It is important to note that anomalous transport may also be observable in certain condensed matter systems, such as the ---so far theoretical--- constructions called the Weyl semi-metals, that can be viewed as strongly coupled electron-hole plasmas in 3 spatial dimensions with the single-particle excitations being chiral fermions \cite{Weyl1,Weyl2,Weyl3}.   

In this paper we address the question whether the anomalous conductivities receive radiative corrections. We address this question in the holographic setting with as much generality as possible.  The axial current is known to enjoy both electromagnetic and QCD quantum anomalies that lead to the anomaly equation, 
\be\lab{aneq}
\6_\m J_5^\m = a_1 F^A \wedge F^A + a_2 F^V \wedge F^V + a_3  \tr\, G\wedge G  \ , 
\ee
where $F^{A,V}$ are the field strengths of background axial\footnote{Even though there is no background axial gauge fields in nature, here we include them for generality.} and vector gauge fields, $G$ is field strength of the gluon field and $a_i$ are anomaly coefficients that are well-known to be one-loop exact \cite{ABJ}. 

In the absence of perpendicular external (axio-)electric and (axio-)magnetic fields the first two terms in (\ref{aneq}) vanish. The last term in (\ref{aneq}) then is the main source of anomaly induced chiral imbalance. One can then effectively take into account such anomaly generating glue transitions\footnote{In QCD at finite temperature the most dominating such process is argued to be the  the sphaleron decays \cite{spdecay}} by introducing an axial chemical potential $\mu_5$. 

Then, one can show by various different methods \cite{CME1,CME2}, that the presence of the electromagnetic anomaly $a_2\neq 0$ in the presence of an external magnetic field  $\vec{B}$  leads to generation of an electric current as in \eqref{J1}, with
\be\lab{sigma1} 
 \sigma_\text{VV} =  \frac{e^2}{2\pi^2} \, \mu_5\, .
\ee 
This result makes explicit use of the anomaly for a single chiral species \cite{ABJ}, and $\sigma_\text{VV}$  is the \emph{chiral magnetic conductivity}. There is a similar effect, namely generation of a chiral current $J^5$ in response 
to a magnetic field that is $J^5 = \sigma_\text{AV} B$. This effect is called the {\em chiral separation effect}. 

\vspace{10pt}
A great deal of theoretical research in anomalous transport is focused on whether the value of the chiral magnetic conductivity  above  (and similarly other anomalous conductivities) receives radiative corrections or not. There exist a variety of arguments  in favor of ---at least perturbative--- non-renormalization \cite{NonRenorm, CME2, JensenNR, Minwalla, Roy, Safodyev1, Safodyev2}, mainly due to the fact that the anomaly coefficients are one-loop exact \cite{ABJ}. The situation 
however is subtle and to make a clear statement about non-renormalization one has to distinguish the two types of anomalies that we will call type I and type II \cite{Kovtun,Banerjee2}. The former type refers to anomalies that vanish when the external fields are turned off (i.e. they are  't Hooft anomalies), such as the first and the second terms in (\ref{aneq}), whereas the type II anomalies refer to mixed gauge-global anomalies such as the gluonic contribution in (\ref{aneq}) whose presence does not depend on external fields. In the latter case, one {\em does} expect radiative corrections generically \cite{Kovtun}.

 In the absence of type II anomalies, on the other hand, there exists both direct and indirect methods establishing non-renormalization. Firstly, one can show  absence of perturbative corrections directly in field theory  using the axial and vector Ward identities and 
some recently proven non-renormalization theorems \cite{NonRenTheorem}, see  for example \cite{Buividovich}.  Absence of non-perturbative renormalization can also be established by two indirect  methods. Firstly, assuming that a hydrodynamic description of the system is at hand, demanding a positive definite divergence of the entropy current determines $\sigma_\text{VV}$ to be exactly as in (\ref{J1}), as shown in \cite{SonSurowka} (see also \cite{NeimanOz}). Secondly, \cite{Yarom} established a Euclidean effective field theory for anomalous transport, whose consistency again requires fixing the value of $\sigma_\text{VV}$ as in (\ref{J1}). 

\vspace{10pt}
In this note we address the question of whether the value of anomalous conductivities such as $\sigma_\text{VV}$ in (\ref{J1}) are exact in  strongly coupled quantum field theories that admit a gravitational dual description a la AdS/CFT \cite{Malda,Polyakov,Witten1}.   
The study of anomalous transport via the holographic correspondence played a major role in the development of the subject from early on \cite{Erdmenger, Banerjee}. In holography one introduces anomalous currents by considering bulk gauge fields in the presence of bulk Chern-Simons terms \cite{Witten1}. One can then calculate the anomalous conductivities using the Kubo's formulae by calculating the retarded Green's functions following the standard prescription of the holographic correspondence. 
The  case of conformal plasma of ${\cal N}=4$ super Yang-Mills in the large $N_c$ limit was first to be considered in the holographic description, in a series of papers by Landsteiner et al. \cite{Landsteiner1,Landsteiner2,Landsteiner3,Landsteiner4}. By comparison of the holographic and weak coupling results, these authors  concluded that the chiral magnetic conductivity receives no corrections at all. 
However, this is a very special theory, and one is immediately prompted to analyze the situation in a more general class of theories, in particular theories with a mass gap and running gauge coupling. Such a study was undertaken very recently by one of the authors together with A. Jansen in \cite{GJ}. In that paper the anomalous conductivities were calculated in a holographic setting that is dual to a non-conformal theory that exhibits a confinement-deconfinement transition. One finds that the value in (\ref{J1}) non-trivially depends on the parameters of the gravitational background, hence it seems that the universal value in (\ref{J1}) no longer holds. However, when the result is expressed in terms of  physical quantities such as the chemical potential $\mu_5$ and temperature $T$, one again finds that the anomalous conductivities  attain their universal values. In particular the chiral magnetic conductivity is again precisely given by (\ref{J1}) \cite{GJ}. 

The non-trivial result obtained in \cite{GJ} prompted us to seek for a {\em generic background-independent mechanism} to explain the universality of anomalous conductivities in the holographic setting. In a sense, in this paper we seek for the {\em holographic analog of the non-renormalization theorems in field theory}, that we summarized above. Such universal values for the transport coefficients would typically result from the universal near-horizon behavior of bulk fluctuations in black-hole backgrounds. The most famous example of such behavior is the universal value of the shear viscosity to entropy density ratio $\eta/s= 1/4\pi$ in gravitational theories quadratic in derivatives \cite{Policastro, BJ}. This robust result can indeed be explained by the background-independence of metric fluctuations near the horizon, see for example \cite{IL}. 

In case of the anomalous conductivities, such direct proofs of universality prove difficult because, unlike the case of the shear viscosity or electric conductivity, calculation of anomalous conductivities in holography involves mixing of bulk gauge and metric fluctuations. In a way, in order to establish universality one has to diagonalize these fluctuations which turns out to be an onerous task. A simpler case was studied in \cite{Landsteiner:2012dm}, where the authors considered gravity theories with no scalars and  looked at the holographic flow of the chiral condutivities, i.e. their  dependence on the radial coordinate. Particularly, in the case of the AdS-Reissner-Nordstr\"om blackhole, they found no holographic flow except the scale dependence of the chemical potentials.    

In this paper, we address the calculation in a general class of two-derivative gravity models in an alternative way, namely by including the source, \emph{i.e.}, the axial or the vector magnetic field,  in the fluctuations themselves. 
This method was already introduced  by Donos and Gauntlett  in a different context where the authors study the thermoelectric properties of holographic plasmas \cite{DG}.  Employing this method we prove the universality of anomalous conductivities such as the chiral magnetic and chiral separation conductivities for a quite general action.  This result establishes the holographic analog of the non-renormalization theorems that are found on the field theory side. 

\vspace{10pt}

The paper is organized as follows. 

In section 2, we introduce the general holographic setting that we employ in this paper. Here we also fix the coefficients of the Chern-Simons terms by matching the anomaly equation on the field theory side. 

In section 3 we introduce our ansatz for the fluctuation equations in order to calculate the anomalous conductivities. We study the near boundary and near horizon behavior of these fluctuations  and show that the regularity of metric fluctuations near the horizon require vanishing of metric fluctuations there. In this section we also derive expressions for the conserved fluxes that correspond to these fluctuations.  

In section 4 we finally evaluate the anomalous conductivities associated with background vector and axial sources and demonstrate universality in their values.  

The final section discusses our results and the methods and present an outlook for further research.

\section{Action and equations of motion}

The model we work with in this paper is given by the following action
\begin{align}
S & = \frac{1}{16\, \pi\,G} \int \Bigg[ R *1 - \frac{1}{2} \d\phi\wedge*\d \phi - \frac{\Psi(\phi)}{2}\d \chi \wedge * \d \chi - V(\phi) * 1 - \frac{Z_A(\phi)}{2} F^A \wedge * F^A \nonumber\\
& \qquad\qquad \qquad - \frac{Z_V(\phi)}{2} F^V \wedge * F^V + \frac{\kappa}{3} \, A \wedge \left( F^A \wedge F^A + 3 \,g\, F^V \wedge F^V \right) \Bigg] \lab{act}  \ , 
\end{align}
with $F^A=\d A$ the axial field and $F^V=\d V$ the vector one. The dilaton scalar field, $\phi$, and the axion one, $\chi$, will not play an explicit r\^ole in the solution of the fluctuations, even when we will assume that they are not-trivial in the background.
The main effect of the dilaton in this study comes through the dilatonic couplings $Z_{A,V}$. Here we assume that the one-forms $A$ and $V$ are normalized in such a way that
\be\label{Zsnormalization}
\lim_{r\to\infty} Z_A (\phi)= \lim_{r\to\infty} Z_V (\phi)=1 \ ,
\ee
with $r\to\infty$ corresponding to the boundary, where the metric is asymptotically AdS$_5$. The normalization \eqref{Zsnormalization} can always be attained with a redefinition of the factors $\kappa$ and $g$ in the Chern-Simons (CS) terms.

The choice of  CS terms in (\ref{act}) corresponds to including a Bardeen counter-term in the boundary action \cite{Landsteiner1} whose presence is required for an anomaly-free vector current. 
Indeed if we make a gauge transformation $V\to V + \d \zeta_{V}$ in (\ref{act}) we obtain $\6_\mu J^\mu=0$ for the vector current that is dual to the bulk gauge-field $V^\mu$.  

The coefficients $\kappa$ and $g$ are not arbitrary, and their value can be found matching to the gauge anomaly of one  left-handed and one right-handed fermion. With a gauge transformation $\delta_{\zeta_A}$ for the axial U(1) field $A$ we obtain 
\begin{equation}\lab{trans}
\delta_{\zeta_A} S = \frac{\kappa}{48\,\pi\,G} \le(F^A\wedge F^A + 3\,g\, F^V\wedge F^V\ri) = - \partial_\mu J_5^\mu\, . 
\end{equation}
On the other hand, in the presence of the Bardeen counter-term on the boundary theory, the correct anomaly equation for the axial current reads, 
\be\lab{axanom} 
 \partial_\mu J_5^\mu =  \frac{1}{12\pi^2} \le( 3 F^V \wedge F^V + F^A \wedge F^A\ri)\, .
 \ee 
Matching (\ref{trans}) with (\ref{axanom}) we determine
\begin{equation}\label{eq.kappavalue}
\kappa = - \frac{4\,G\, N_c}{\pi}, \qquad g=1\ .
\end{equation}
From now on we set $g=1$.

\vspace{10pt}

The equations of motion from the variation of the action \eqref{act} read for the scalars
\begin{align}
\d \left( \Psi \, *\d\chi \right) & = 0 \ , \\
\d  *\d\phi & = \partial_\phi V\, *1 + \frac{\partial_\phi Z_A}{2}\, F^A \wedge *F^A + \frac{\partial_\phi Z_V}{2}\, F^V \wedge *F^V + \frac{\partial_\phi \Psi}{2}\, \d \chi \wedge * \d \chi  \ ,
\end{align}
for the gauge fields 
\begin{align}\label{eq.U1field}
\d \left( Z_A\, *F^A - \kappa \, A \wedge F^A - \kappa \,  V \wedge F^V  \right) & = 0 \ , \\ \label{eq.U1field2}
\d \left( Z_V\, *F^V - 2 \kappa \,  A \wedge F^V  \right) & = 0  \ ,
\end{align}
and finally for the metric
\begin{align}
R_{\mu\nu} & = \frac{1}{2}\partial_\mu \phi \partial_\nu \phi + \frac{\Psi}{2} \partial_\mu \chi \partial_\nu \chi +\frac{V}{3} g_{\mu\nu} + \frac{Z_A}{2} \left( F^A_{\mu\rho}{F^A_\nu}^\rho - \frac{1}{6} g_{\mu\nu} F^A_{\rho\sigma} F^{A,\rho\sigma} \right) \nonumber\\
& \quad + \frac{Z_V}{2} \left( F^V_{\mu\rho}{F^V_\nu}^\rho - \frac{1}{6} g_{\mu\nu} F^V_{\rho\sigma} F^{V,\rho\sigma} \right) \ .\label{einstein}
\end{align}

\vspace{10pt}
We will assume a static, translation- and rotation-invariant  background, given by the following configuration
\begin{align}\label{eq.bgansatz}
\d s^2 & = - g_{tt}(r) \d t^2 + g_{xx}(r) \d \vec x^2 + g_{rr}(r) \d r^2 \ , \\
A & = A_t(r) \d t \ , \qquad  V  = V_t(r) \d t \ , \qquad  \phi(r) \ , \qquad \chi(r) \ .
\end{align}
Notice that $g_{tt}$ is a positive-definite function, and that AdS-RN with two charges falls into this general ansatz for the specific values  $Z_A=Z_V=1$, $V=-20$, $\chi=\phi=0$. If we further require $V_t=0$ we recover the case studied in section 3.2 of \cite{Landsteiner:2012dm}.

In the UV, $r\to\infty$, we will require that the solution becomes asymptotically AdS$_5$
\begin{equation}\label{eq.UVbg}
g_{tt} \sim r^2 + \cdots \ , \qquad
g_{xx}  \sim r^2 + \cdots \ ,   \qquad
g_{rr} \sim r^{-2} + \cdots \ , 
\end{equation}
 with the remaining functions going to constants: $A_t \sim A_t^\infty+\cdots$, etc.
The dots indicate subleading  terms.

For the IR of the theory we require the existence of a non-extremal horizon, such that near the horizon, $r= r_h$, we have
\begin{equation}\label{eq.IRbg}
g_{tt}  \sim t_h (r-r_h) + \cdots \ , \qquad
g_{xx}  \sim x_h + \cdots \ ,  \qquad
g_{rr}  \sim \frac{\rho_h}{r-r_h} + \cdots \ , 
\end{equation}
with the remaining functions going to constants: $A_t \sim A_t^h+\cdots$, etc.

Before analysing the fluctuations of the system it is useful to express the temporal components of the gauge fields, $A_t$ and $V_t$, in terms of constants of motion. To this end let us define first
\begin{align}\label{eq.Jform}
J_5^{\mu\nu} & =   - \sqrt{-g}\, Z_A(\phi) F^{A,\mu\nu} + \frac{\kappa}{2}\, \epsilon^{\mu\nu\alpha\rho\sigma} \left(  A_{\alpha}F^A_{\rho\sigma} +  a_\alpha F^V_{\rho\sigma} \right) \ , \\
J^{\mu\nu} &  =  - \sqrt{-g} \, Z_V(\phi) F^{V,\mu\nu} + \kappa  \,\epsilon^{\mu\nu\alpha\rho\sigma}   A_{\alpha}F^V_{\rho\sigma} \ ,\label{eq.Jform2}
\end{align}
from where the equations of motion \eqref{eq.U1field} and \eqref{eq.U1field2} read simply
\begin{equation}
\partial_\mu J_5^{\mu\nu} = 0 \ , \qquad  \partial_\mu J^{\mu\nu} = 0 \ .
\end{equation}
For the background \eqref{eq.bgansatz} the only  non-vanishing components are
\begin{equation}\label{eq.chargedensity}
J_{5,bg}^{rt} = \sqrt{\frac{g_{xx}^3}{g_{tt}g_{rr}}} Z_A(\phi)\, A_t' \ , \qquad J_{bg}^{rt} = \sqrt{\frac{g_{xx}^3}{g_{tt}g_{rr}}} Z_V(\phi)\, V_t' \ ,
\end{equation}
which from the equations of motion are constants
\begin{equation}
J^{rt}_{5,bg}=Q_5 \ , \qquad J^{rt}_{bg}=Q \ .
\end{equation}

\section{Fluctuations}

In this section we give the configuration of fluctuations that will give rise to the anomalous transport coefficients. As shown for example in \cite{Landsteiner1}, it is consistent to restrict the study to the following set of fluctuations
\begin{align}
\delta A & = \delta A_x(y,r)\d x + \delta A_z(y,r) \d z \ , \\
\delta V & = \delta V_x(y,r)\d x + \delta V_z(y,r) \d z \ , \\
\delta \d s^2 & = 2 \,\delta g_{tx}(y,r) \d t \,\d x + 2\, \delta g_{tz}(y,r) \d t \,\d z \ .
\end{align}
These fluctuations correspond to the vectorial sector preserving rotations in the $x$--$z$ plane, which is the reason why they do not couple to fluctuations of other components of the metric or the gauge fields, nor to the scalars.

In this moment we will employ the method in \cite{DG} to express the sources explicitely in the fluctuations. We will turn these on for the gauge fields only in the following way
\begin{align}\label{eq.ansatz}
\delta A_x (y,r) & = -B^5_z \,y + \alpha_x(r) \ , \qquad
\delta A_z (y,r)  = B^5_x \,y + \alpha_z(r) \ , \\
\delta V_x (y,r) & = -B_z \,y + \beta_x(r) \ , \qquad
\delta V_z (y,r)  = B_x \,y + \beta_z(r) \ , \\
\delta g_{tx} (y,r) & = g_{xx} \, \gamma_x(r) \ , \qquad\qquad\,\,
\delta g_{tz} (y,r)  = g_{xx} \, \gamma_z(r) \ ,
\end{align}
with the magnetic field sources $B_a$ and $B_a^5$, with $a=\{x,z\}$.
Now, we want the $\alpha_a$ and $\gamma_a$ to correspond to normalizable deformations of the fields, thus describing the response to the $B_a^{(5)}$ sources. Since the gauge fields and the graviton are massless, the holographic correspondence implies that  we must have near the UV the following leading behavior:
$
\alpha_a \sim r^{-2} 
$, 
$
\beta_a \sim r^{-2}
$ and
$
\gamma_a \sim r^{-4}
$.

To check this asymptotic behavior let us begin analyzing the  equations of motion for the gauge fields fluctuations. With the definitions \eqref{eq.Jform} and \eqref{eq.Jform2} these equations are
\begin{equation}
\partial_r \, \delta J_5^{r \,a} + \partial_y \, \delta J_5^{y\,a} = 0 \ , \qquad \partial_r \, \delta J^{r \,a} + \partial_y \, \delta J^{y\,a} = 0 \ ,
\end{equation}
with $\delta J_{(5)}^{\mu\,a}$ the part of \eqref{eq.Jform2} (\eqref{eq.Jform}) linear in fluctuations with our background ansatz \eqref{eq.bgansatz}.

It turns out that the derivatives with respect to the spatial coordinate $y$  can be expressed as radial derivatives
\begin{equation}
\partial_y \, \delta J_5^{y\,a}  = - \left( \kappa\, A_t' \, B^5_b + \kappa\, V_t'\, B_b \right) \delta^{ab} \ , \qquad
\partial_y \, \delta J^{y\,a}  = - 2 \kappa \,V_t' \, B^5_b\, \delta^{ab} \ ,
\end{equation}
with $\delta^{ab}$ Kronecker's delta.
Therefore the equations of motion for the fluctuations of the gauge fields read 
\begin{equation}\label{eq.constantofmotion}
\partial_r \tilde J_5^a = \partial_r \tilde J^a = 0 \ ,
\end{equation}
where we have defined
\begin{align}\label{eq.Jconstants}
\tilde J_5^a & \equiv  -\left( 2\kappa\, A_t\, B^5_b+ 2\kappa\, V_t\, B_b + Q_5\, \gamma_b + \sqrt{\frac{g_{tt}g_{xx}}{g_{rr}}} Z_A(\phi) \, \alpha_b'  \right) \delta^{ab}   \ , \\
\tilde J^a & \equiv -\left(2 \kappa \, A_t\, B_b + 2 \kappa \,V_t\, B^5_b + Q\, \gamma_b + \sqrt{\frac{g_{tt}g_{xx}}{g_{rr}}} Z_V(\phi) \, \beta_b'  \right) \delta^{ab}\ .\label{eq.Jconstants2}
\end{align}

The $\tilde J^a_{(5)}$ quantities are conserved on-shell along the radial direction. In particular they help determining the leading behavior of the $\alpha_a$ and $\beta_a$ fluctuations in the UV. Provided $\gamma_a\sim r^{-4}$ we indeed get
\begin{align}\label{eq.alphaUV}
\alpha_a(r) & \simeq \frac{\tilde J_5^{b} \delta_{ab} + 2\kappa\left( A_t^{\infty}\,B^5_a  + V_t^{\infty}\,B_a \right)}{2\,r^2} + \cdots \ ,  \\
\beta_a(r) & \simeq \frac{\tilde J^{b} \delta_{ab} + 2\kappa \left( A_t^{\infty}\,B_a  +  V_t^{\infty}\,B^5_a \right)}{2\,r^2} + \cdots \ , \label{eq.betaUV}
\end{align}
at large radius, and notice the shift with respect to the conserved quantities $\tilde J_5^a$ in the numerator.

We must now prove that consistently $\gamma_a\sim r^{-4}$ near the boundary. This is straightforward to see once we realize that the equations of motion for the fluctuations $\gamma_a$ are simply
\begin{equation}
 \partial_r \tilde K_{a} = 0\ ,
\end{equation}
where we have used the background equations of motion and defined two more constants of motion
\begin{equation}\label{eq.Kconstants}
\tilde K_a \equiv \sqrt{ \frac {g_ {xx}^5} {g_ {tt} g_ {rr}}} \gamma_a'  + Q_5 \, \alpha_a  + Q \, \beta_a \ .
\end{equation}

Evaluating $\tilde K_a$ at the UV and using the boundary behavior of the background solutions we find at large radius
\begin{equation}\label{eq.gammaUV}
\gamma_\mu(r) \simeq - \frac{\tilde K_\mu}{4\, r^4}  + \cdots \ , 
\end{equation}
and we recover from the fluctuation equations the expected behavior for the functions. 

\vspace{10pt}

We  study now the behavior of the fields near the horizon. With this in mind let us first quote the result for the scalar of curvature with the fluctuations of the metric \eqref{eq.ansatz}:
\begin{equation}
R = R_{bg} + \frac{{\cal S}^a\, \gamma_a+ {\cal S}^{ab}\, \gamma_a \gamma_b  }{g_{tt}} \ ,
\end{equation}
where ${\cal S}^a$ and ${\cal S}_{ab}$ are some radial functions \emph{regular} at the horizon, and $R_{bg}$ the Ricci scalar  of the background solution. From this expression we see that near the horizon one has a curvature singularity unless the $\gamma_a$ vanish there.

Plugging this behavior in \eqref{eq.Jconstants}, \eqref{eq.Jconstants2} and \eqref{eq.Kconstants} the leading behavior of $\alpha_a$ compatible with the equations is that the fluctuations of the gauge fields go to constants at the horizon.

With these two results in hand we can determine the value of the constants of motion $\tilde J^a_{(5)}$ straightforwardly at the horizon
\begin{align}\label{eq.tildeJvaluehorizon}
\tilde J_5^a = -2\left( \kappa\, A_t^h\, B^5_b + \kappa\, V_t^h B_b \right) \delta^{ab} \ , \\
\tilde J^a = -2\, \kappa \left( A_t^h\, B_b + V_t^h B^5_b \right) \delta^{ab} \ .\label{eq.tildeJvaluehorizon2}
\end{align}

Once we have determined the behavior of the fields near the horizon and near the boundary we have specified completely the solutions to $\alpha_a$, $\beta_a$ and $\gamma_a$. In the next section we will see how this is enough to build the one-point functions associated to the fluctuations we have been considering.

\section{Chiral magnetic and separation effects}

Once equipped with the expressions \eqref{eq.tildeJvaluehorizon} and \eqref{eq.tildeJvaluehorizon2} we can write the one-point functions corresponding to the expected value of the axial and vector currents. This is obtained from the normalizable mode of the $\alpha_a$ (in equation \eqref{eq.alphaUV}) and $\beta_a$ (in equation \eqref{eq.betaUV}) fluctuations appropriately normalized.
From  action \eqref{act} the final result reads\footnote{Strictly speaking one should choose the gauge $A_t^\infty = V_t^\infty = 0$ to avoid complications in the calculation by a redefinition of \eqref{J5final}, see \cite{LandsteinerR}. We discuss this point further in the Discussion section below.}
\begin{align}\label{J5final}
\langle J_5^a \rangle =  \frac{- \kappa}{8\pi\,G} \left[ (A_t^\infty-A_t^h) \, B^5_b + (V_t^\infty-V_t^h) \, B_b \right] \delta^{ab}=  \frac{- \kappa}{8\pi\,G}\left(\mu_5\, B^5_b +  \mu \, B_b \right)\delta^{ab}\ , \\
\langle J^a \rangle =  \frac{- \kappa}{8\pi\,G} \left[ (A_t^\infty-A_t^h) \, B_b + (V_t^\infty-V_t^h) \, B^5_b \right]\delta^{ab} =  \frac{- \kappa}{8\pi\,G}\left( \mu_5\, B_b +  \mu\, B^5_b \right)\delta^{ab}\ ,
\end{align}
with  $A_t^\infty-A_t^h=\mu_5$ and $V_t^\infty-V_t^h=\mu$ the associated chemical potentials. Plugging in the value for $\kappa$ in \eqref{eq.kappavalue} we get for the one-point functions
\begin{align}
\langle J_5^a \rangle = \frac{ N_c}{2\pi^2}\left(\mu_5\, B^5_b + \mu \, B_b \right)\delta^{ab}\ , \\
\langle J^a \rangle =  \frac{ N_c}{2\pi^2}\left( \mu_5\, B_b +  \mu\, B^5_b \right)\delta^{ab}\ .
\end{align}
From these expressions we define the chiral conductivities as the derivatives of the one-point function with respect to each of the magnetic fields. Since the dependence on these is linear we can simply write
\begin{align}
\langle J^a \rangle  & = \sigma_\text{VV} B^a +  \sigma_\text{VA} B^{5\,a} \ , \\ 
\langle J_5^a \rangle  & =  \sigma_\text{AA} B^{5\,a}+ \sigma_\text{AV} B^a  \ , 
\end{align}
with
\begin{equation}
\sigma_\text{VV} = \sigma_\text{AA} = \frac{N_c}{2\pi^2}  \mu_5 \ , \qquad \sigma_\text{AV} = \sigma_\text{VA} = \frac{N_c}{2\pi^2}  \mu \ .
\end{equation}

\vspace{10pt}
\noindent
In particular, upon setting the axial magnetic field to zero, we obtain the desired {\em chiral magnetic effect}
\be\lab{CME}
\la J^\nu \ra = \sigma_\text{VV} B^\nu  = \frac{N_c}{2\pi^2}  \mu_5 B^\nu\, ,
\ee
and  {\em chiral separation effect}
\be\lab{CSE}
\la J^\nu_5 \ra = \sigma_\text{AV} B^\nu  = \frac{N_c}{2\pi^2}  \mu \,B^\nu\, .
\ee
We emphasize that our results (\ref{CME}) and (\ref{CSE}) follows directly from the horizon universality of the fluctuations and, as such, they are valid in a generic class of theories where the gravitational description solves the generic action (\ref{act}).  

\section{Discussion} 

We demonstrated that the anomalous conductivity describing the chiral magnetic effect acquires an universal value for a generic holographic model, independently of the details of the background solution on which this effect is calculated. As we discussed in the Introduction, this fact was already shown to hold on the field theory side with a variety of different methods including the Ward identities and new non-renormalization theorems, hydrodynamics and effective field theories. Our calculation provides yet another, independent demonstration of this non-renormalization and fills in the gap on the dual gravitational side. It is reassuring to find that there exist a holographic analog to this field theory non-renormalization theorem, and it indeed follows from horizon universality, as one would have expected.   

The only requirement that we impose in our construction, besides the general form of the action, is a rather physical one: that the curvature scalar is not divergent on the horizon. This allowed us to express the conserved quantities \eqref{eq.Jconstants}, \eqref{eq.Jconstants2} and \eqref{eq.Kconstants} in terms of horizon quantities, which recombined in a neat way with UV data in the solution for the fluctuations near the boundary to produce the chemical potentials of the theory in the final result.

We left out the calculation of the chiral vortical effect in our analysis. This can be studied by adding an extra piece to the action \eqref{act}
\begin{equation}
S \to S + \frac{\lambda}{16\pi G} \int A \wedge \text{tr}\left( {\cal R}\wedge {\cal R} \right) \ ,
\end{equation}
with ${\cal R}^\mu{_\nu} = R^\mu{_{\nu\rho\sigma}} \d x^\rho \wedge \d x^\sigma$ the curvature tensor. Additionally one needs to add a new counterterm to the action to make the variational problem well posed. The effect of this new piece of the action in the equations of motion is to add an extra piece that behaves as $ \lambda\, \text{tr} ({\cal R}\wedge{\cal R})$,  in \eqref{eq.U1field}, and a new one in \eqref{einstein} as well. For the fluctuations considered in this paper the former term vanishes, impliying that the result for the anomalous conductivity cannot change due to the presence of the new term in the action. We plan to present the calculation for all anomalous transport coefficients including the chiral vortical effect in the future, thus providing the generalization for a gravity theory with scalar matter of the results in section 4 of \cite{Landsteiner:2012dm}.

We should also comment on a technical issue that was alluded to in the footnote on the previous page. Our result should of course be independent of the choice of gauge for the bulk gauge fields. However, there is a subtlety \cite{LandsteinerR} arising from different methods to realize the chemical potentials $\mu$ and $\mu_5$ on gravity side. In the first method, that we employed here, we choose a vanishing value for the boundary values of the gauge fields. In this method the calculation goes straightforwardly with the definition \eqref{J5final}.  One may also choose the gauge fields to vanish on the horizon and asymptote to finite values on the boundary. In this case however  there is another contribution to the boundary chiral current \eqref{J5final} that arises from the finite Chern-Simons current that is added to the {\em consistent} current to obtain a {\em covariant} current. To establish gauge equivalence with the previous method one has to include a spurious boundary axion \cite{LandsteinerR}.  We choose to work with the first method, i.e. the formalism B in the nomenclature of \cite{LandsteinerR}. 

\vspace{10pt}

As is common with general calculations showing a robust result, the setup we have considered in this paper may help identify ways to model setups where the chiral magnetic conductivity differs from the universal result found here. 
One such possibility is to try to prevent the equation of motion for the axial gauge field fluctuation to have a constant of motion, \eqref{eq.constantofmotion}. This follows for example from a St\"uckelberg type of action, in which the kinetic term for the axion in \eqref{act} now reads
\begin{equation}
(16\pi G)\,S_{\chi\chi} = - \int \frac{\Psi(\phi)}{2} \left( \d \chi - m_A A \right) \wedge * \left( \d \chi - m_A A \right)  \ ,
\end{equation} 
with $m_A$ the mass of the axial field. Indeed this mechanism was proposed as a holographic dual to anomalous theories with type II anomalies ---that is, with a gluonic contribution to the anomaly equation ($a_3\neq 0$ in equation (\ref{aneq}))--- in \cite{GJ}, see also  \cite{LandsteinerII}.  The bulk axion is dual to a theta-term in the field theory action. Coupling $\chi$ to the axial field induces a non-trivial value for the $\langle \text{Tr}\,G\wedge G \rangle$ expectation value in \eqref{aneq}. The anomaly term associated with this expectation value, $a_3$, then induces a modification in the result for the anomalous transport coefficients \cite{GJ,LandsteinerII}. In holographic constructions of QCD $m_A$ is usually proportional to the number of flavors in the theory, and  one needs a calculation in the Veneziano limit to see the effect of this term \cite{workinprogress}. 

There are possible various extensions of our work. First of all, as already mentioned above, the extension of the holographic non-renormalization to the case of chiral vortical conductivity would be very interesting. Secondly, one may wonder whether our universal result for  the chiral conductivities survive the higher derivative corrections in gravity, or not. It is well-known that one generates corrections to the shear viscosity in presence of higher derivative corrections \cite{Brigante}, but the notion of horizon universality continues to hold. In the case of anomalous transport we also expect horizon universality to determine the values of conductivities exactly, also in presence of higher derivative terms. Whether the actual value changes or not remains to be seen. In this context one should note  a physical distinction between viscosity and anomalous conductivities, namely the former is dissipative and the latter is not. It is conceivable therefore  that the result we found here may be robust against higher derivative corrections. Finally, one may wonder if the kind of universality we find here extends to finite frequency and momenta. It would be also very interesting to predict such universal behavior in the momenta dependence of the anomalous conductivities in holography.  

\section*{Acknowledgements}
We thank Francisco Pena-Benitez, Aron Jansen and especially Aristomenis Donos for interesting discussions and useful remarks. JT is grateful to the Mainz Institute for Theoretical Physics (MITP) for its hospitality and partial support during the initial stage of this work. 

This work is part of the D-ITP consortium, a program of the Netherlands Organisation for Scientific Research (NWO) that is funded by the Dutch Ministry of Education, Culture and Science (OCW). 

JT is supported by grants 2014-SGR-1474, MEC FPA2010-20807-C02-01, MEC FPA2010-20807-C02-02, ERC Starting Grant HoloLHC-306605 and by the Juan de la Cierva program of the Spanish Ministry of Economy.

\end{document}